%% file: og.tex
\documentclass[sigplan,screen]{acmart}
\copyrightyear{2020}
\acmYear{2020}
\setcopyright{rightsretained}
\acmPrice{}
\acmConference[CGO '20]{Proceedings of the 18th ACM/IEEE International Symposium on Code Generation and Optimization}{February 22--26, 2020}{San Diego, CA, USA}
\acmBooktitle{Proceedings of the 18th ACM/IEEE International Symposium on Code Generation and Optimization (CGO '20), February 22--26, 2020, San Diego, CA, USA}
\acmDOI{10.1145/3368826.3377909}
\acmISBN{978-1-4503-7047-9/20/02}

\usepackage{booktabs}   
\usepackage{subcaption} 

\usepackage{graphicx}
\usepackage{balance}  

\usepackage{soul}
\usepackage{booktabs}
\usepackage{subcaption}
\usepackage{xspace}
\usepackage[normalem]{ulem}
\usepackage{listings}%
\usepackage{xcolor}
\usepackage[font=small,labelfont=bf,aboveskip=2pt,belowskip=0pt]{caption}
\usepackage{parcolumns}
\usepackage{algorithm}
\usepackage[noend]{algpseudocode}
\usepackage{enumitem}

\usepackage{bm}
\usepackage{lstlinebgrd}
\usepackage{arydshln}

\newcommand{\myparagraph}[1]{\vspace{1.5pt} \noindent {\bf #1.}}
\newcommand\punt[1]{}

\newcommand{\graphit}{GraphIt\xspace}

\newcommand{\saman}[1]{{\color{brown} {\bf Saman:} #1}}

\definecolor{ao(english)}{rgb}{0.0, 0.5, 0.0}

\newcommand{\deltastepping}{$\Delta$-stepping}
\newcommand{\astar}{A$^{*}$ search}
\newcommand{\kcore}{$k$-core}

\newcommand{\scheduleKeyword}{{\small\texttt{schedule}\xspace}}

\newcommand{\Fig}{Fig.}

\newcommand{\DensePull}{DensePull\xspace}

\newcommand{\PriorityCoarsening}{Priority coarsening\xspace}
\newcommand{\priorityCoarsening}{priority coarsening\xspace}
\newcommand{\OG}{GraphIt\xspace}
\newcommand{\extension}{new priority-based extension\xspace}

\newcommand{\stt}[1]{{\small \texttt{#1}}}
\newcommand{\ftt}[1]{{\footnotesize \texttt{#1}}}

\newenvironment{denseitemize}{
\begin{itemize} [topsep=2pt, partopsep=0pt, leftmargin=1.5em]
 \setlength{\topsep}{0pt}
 \setlength{\itemsep}{2pt}
 \setlength{\parskip}{0pt}
 \setlength{\parsep}{0pt}
}{\end{itemize}}

\let\origthelstnumber\thelstnumber
\makeatletter
\newcommand*\Suppressnumber{%
  \lst@AddToHook{OnNewLine}{%
    \let\thelstnumber\relax%
     \advance\c@lstnumber-\@ne\relax%
    }%
}

\newcommand*\Reactivatenumber[1]{%
  \setcounter{lstnumber}{\numexpr#1-1\relax}
  \lst@AddToHook{OnNewLine}{%
   \let\thelstnumber\origthelstnumber%
   \refstepcounter{lstnumber}
  }%
}

\newcommand{\maxSpeedup}{3}
\newcommand{\maxSlowdown}{6$\%$}

\newcommand{\numApps}{six\xspace}

\makeatother

\usepackage{microtype}
\usepackage{subcaption,wrapfig}
\usepackage{balance}
\input{graphitdef.tex}

\begin{document}

\title[Optimizing Ordered Graph Algorithms with \OG{}]
{Optimizing Ordered Graph Algorithms with \OG{}}

\author{Yunming Zhang}
\affiliation{
  \institution{MIT CSAIL}            
  \country{USA}
}\email{yunming@mit.edu} 

\author{Ajay Brahmakshatriya}
\affiliation{
  \institution{MIT CSAIL}            
  \country{USA}
}\email{ajaybr@mit.edu}

\author{Xinyi Chen}
\affiliation{
  \institution{MIT CSAIL}            
  \country{USA}
}\email{xinyic@mit.edu}

\author{Laxman Dhulipala}
\affiliation{
  \institution{Carnegie Mellon University}            
  \country{USA}
}\email{ldhulipa@cs.cmu.edu}

\author{Shoaib Kamil}
\affiliation{
  \institution{Adobe Research}            
  \country{USA}
}\email{kamil@adobe.com}

\author{Saman Amarasinghe}
\affiliation{
  \institution{MIT CSAIL}            
  \country{USA}
}\email{saman@csail.mit.edu}

\author{Julian Shun}
\affiliation{
  \institution{MIT CSAIL}            
  \country{USA}
}\email{jshun@mit.edu}

\input{abstract}



\maketitle
\renewcommand{\shortauthors}{Y. Zhang, A. Brahmakshatriya, X. Chen, L. Dhulipala, S. Kamil, S. Amarasinghe, J. Shun}

\input{intro}

\input{prelim}
\input{tradeoff}

\input{programming}
\input{compiler}

\input{eval}

\input{related}

\section{Conclusion}
We introduce a new priority-based extension to \graphit that simplifies the programming of parallel ordered graph algorithms and generates high-performance implementations. 
We propose a novel bucket fusion optimization that significantly improves the performance of many ordered graph algorithms on road networks. 
\OG with the extension achieves up to \maxSpeedup$\times$ speedup on \numApps ordered algorithms over state-of-the-art frameworks (Julienne, Galois, and GAPBS) while significantly reducing the number of lines of code.

\input{artifact}
\input{ack}


\balance
\bibliography{graph.bib}  



%

\end{document}

%% file: graphitdef.tex
\definecolor{gray}{gray}{0.5}
\definecolor{key}{rgb}{0,0.5,0}

\def\OPTL{\textrm{$[$}}
\def\OPTR{\textrm{$]$}}

\definecolor{lightbackground}{rgb}{.98,.98,.97}
\definecolor{darkgray}{rgb}{.3,.3,.3}
\definecolor{darkred}{rgb}{.6,0,0}
\definecolor{darkgreen}{rgb}{0,.6,0}
\definecolor{darkblue}{rgb}{0,0,.6}

\lstdefinelanguage{graphit}{%
  keywords={[1]},
  keywords={[2]func,end,element,const,var, field,for,%
    vertexset,edgeset,vector,priority_queue,%
    void,char,short,long,int,float,double,boolean,size_t},
  keywords={[3]filter, from, to, srcFilter, dstFilter, apply, applyModified, applyUpdatePriority},
  keywords={[4],schedule, s1, l1, l2, l3 %
    },
  literate={[OPT[}{{\OPTL}}1 {]OPT]}{{\OPTR}}1,
  string=[b]",
  comment=[l]//,
  morecomment=[s]{/*}{*/},
  mathescape=true,
  flexiblecolumns=true,
  tabsize=2,
  captionpos=b,
  frame=single,
  framerule=0pt,
  aboveskip=2pt,
  belowskip=1pt,
  framesep=1pt,
  basicstyle=\scriptsize\ttfamily,
  keywordstyle={[1]\color{darkred}},
  keywordstyle={[2]\color{blue}},
  keywordstyle={[3]\color{black}\bfseries},
  keywordstyle={[4]\color{darkred}\bfseries},
  numbers=left,
  stepnumber=1,
  numbersep=3pt,
  numberstyle=\tiny,
  emphstyle=\slshape,
  commentstyle=\color{darkgray},
  stringstyle=\color{darkgreen},
  xleftmargin=4.0ex,
}

%% file: abstract.tex
\begin{abstract}

Many graph problems can be solved using ordered parallel graph
algorithms that achieve significant speedup over their unordered
counterparts by reducing redundant work.
This paper introduces a new priority-based extension to \OG{}, a domain-specific language for writing graph applications, to simplify writing high-performance parallel ordered graph algorithms.
The extension enables
vertices to be processed in a dynamic order while hiding low-level implementation
details from the user.
We extend the compiler with new program analyses, transformations, and code generation to produce fast implementations of ordered parallel graph algorithms. 
We also introduce \emph{bucket fusion}, a new performance optimization 
that fuses together different rounds of ordered algorithms to reduce synchronization overhead, resulting in $1.2\times$--\maxSpeedup$\times$ speedup over the fastest existing ordered algorithm implementations on road networks with large diameters.
With the extension, \OG achieves up to \maxSpeedup$\times$ speedup on six
ordered graph algorithms over state-of-the-art frameworks and hand-optimized implementations (Julienne, Galois, and GAPBS) that support ordered algorithms.

\end{abstract}

\punt{Furthermore, our system can automatically discover high-performance optimization strategies using an autotuner built on top of GraphIt. }

\punt{
operators for ordered graph algorithms based on the priorities of vertices.  We implemented the priority-based operators on top of GraphIt, a domain-specific language for graph application that separates the algorithm specification (algorithm language) from the optimization strategy (scheduling language).

  We extend the Graph Iteration Space model, the scheduling language, and the code generator of GraphIt to allow programmers to try out combinations of optimizations designed for the ordered algorithms. We introduce a new bucket fusion optimization that reduces synchronization overheads.

}

%% file: intro.tex
\section{Introduction}
\label{sec:intro}

\begin{figure}[h]
\centering
    \includegraphics [width=0.9\columnwidth] {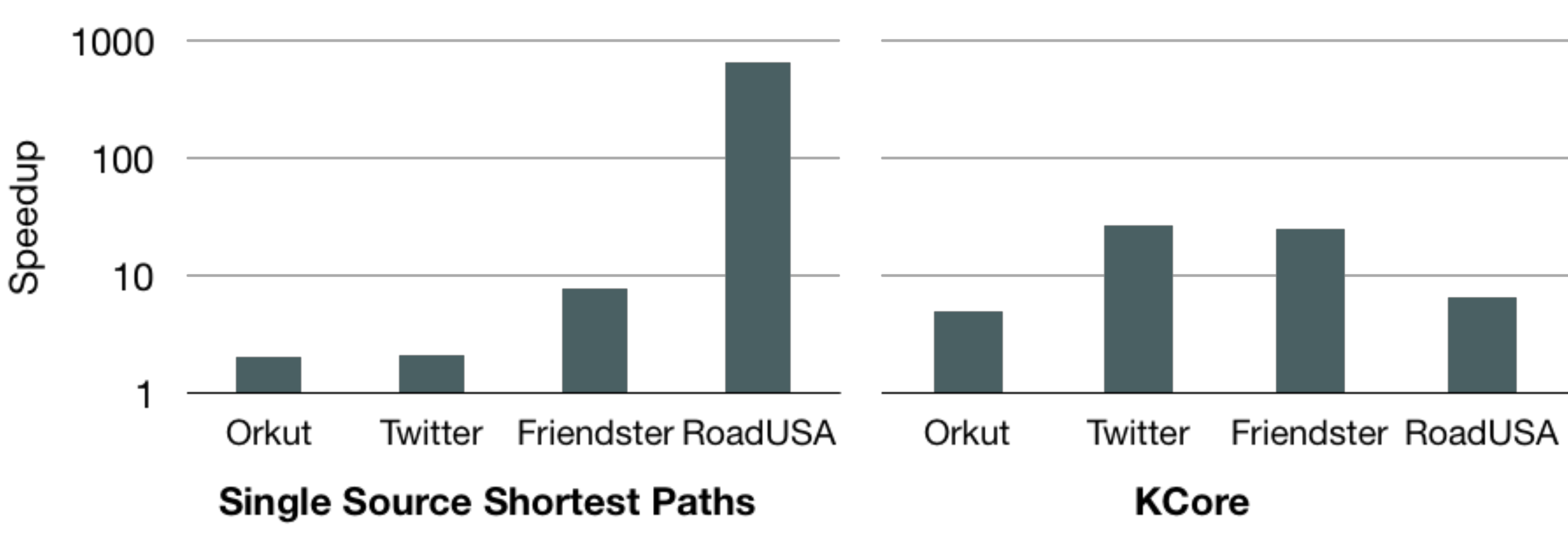}
    \caption{Speedup of ordered algorithms for single-source shortest
    path and $k$-core over the corresponding unordered algorithms implemented in our framework on a 24-core machine.}
    \label{fig:order_vs_unorder_speedup}
\end{figure}

Many important graph problems can be implemented using either \emph{ordered} or \emph{unordered} parallel algorithms.
 \emph{Ordered} algorithms process active vertices following a dynamic priority-based ordering, potentially reducing redundant work. 
By
contrast, \emph{unordered} algorithms process active vertices in an
arbitrary order, improving parallelism while potentially performing a
significant amount of redundant work. In practice, optimized ordered
graph algorithms
are up to two orders of magnitude faster than the unordered
versions~\cite{Dhulipala:2017,Hassaan:2011:OVU:1941553.1941557,DBLP:journals/corr/BeamerAP15,Hassaan:2015:KDG:2694344.2694363}, 
as shown in Figure~\ref{fig:order_vs_unorder_speedup}. 
For example, computing single-source shortest paths (SSSP) on graphs
with non-negative edge weights can be implemented either using the Bellman-Ford
algorithm~\cite{BellmanFord} (an unordered algorithm) or the \deltastepping{}
algorithm~\cite{MEYER2003114} (an ordered algorithm).\footnote{In this paper, we define \deltastepping{} as an ordered algorithm, in contrast to previous work~\cite{Hassaan:2011:OVU:1941553.1941557} which defines \deltastepping{} as an unordered algorithm.}
Bellman-Ford updates the shortest path distances to all active vertices on every iteration.
On the other hand, \deltastepping{} reduces the number of vertices 
that need to be processed every iteration 
by updating path distances to vertices that are
closer to the source vertex first, before processing vertices farther
away.


Writing high-performance ordered graph algorithms is challenging for users who are 
not experts in performance optimization. 
Existing frameworks that support ordered graph algorithms~\cite{Dhulipala:2017,DBLP:journals/corr/BeamerAP15,nguyen13sosp-galois} require users to be familiar with C/C++ data structures, 
atomic synchronizations, bitvector manipulations, and other performance optimizations. 
For example, Figure~\ref{fig:Julienne_delta_stepping} shows a snippet of a user-defined function for \deltastepping{} in Julienne~\cite{Dhulipala:2017}, a state-of-the-art framework for ordered graph algorithms. The code involves atomics and low-level C/C++ operations.


\begin{figure}[t]
\begin{lstlisting} [language=c++,escapechar=/]
constexpr uintE TOP_BIT = ((uintE)INT_E_MAX) + 1;
constexpr uintE VAL_MASK = INT_E_MAX;
struct Visit_F {
  array_imap<uintE> dists;
  Visit_F(array_imap<uintE>& _dists) : dists(_dists) { }
  ....
  inline Maybe<uintE> updateAtomic(uintE& s, uintE& d, intE& w) {
    uintE oval = dists.s[d];
    uintE dist = oval | TOP_BIT;
    uintE n_dist = (dists.s[s] | TOP_BIT) + w;
    if (n_dist < dist) {
      if (!(oval & TOP_BIT) && CAS(&(dists[d]), oval, n_dist)) {
        return Maybe<uintE>(oval);}
      writeMin(&(dists[d]), n_dist);}
    return Maybe<uintE>();}
  inline bool cond(const uintE& d) const { return true; }};
\end{lstlisting}
\caption{Part of Julienne's \deltastepping{} edge update function, corresponding to Lines~\ref{line:udf_start}--\ref{line:udf_end} of \Fig~\ref{fig:code:delta_stepping_GraphIt} in \OG{}'s \deltastepping{}.  }
\label{fig:Julienne_delta_stepping}
\end{figure}

\begin{figure}[t]
\begin{lstlisting} [
	language=graphit,escapechar=|, 
    linebackgroundcolor={%
    \ifnum\value{lstnumber}=5
            \color{purple!10}
    \fi
    \ifnum\value{lstnumber}=9
            \color{purple!10}
    \fi
    \ifnum\value{lstnumber}=15
            \color{purple!10}
    \fi
     \ifnum\value{lstnumber}=16
            \color{purple!10}
    \fi
    \ifnum\value{lstnumber}=18
            \color{purple!10}
    \fi
    \ifnum\value{lstnumber}=19
            \color{purple!10}
    \fi
    }
]
element Vertex end | \label{line:vertexset_edgeset_setup_begin}|
element Edge end
const edges : edgeset{Edge}(Vertex,Vertex, int)=load(argv[1]); |\label{line:vertexset_edgeset_setup_end}|
const dist : vector{Vertex}(int) = INT_MAX; |\label{line:dist_init}|
const pq: priority_queue{Vertex}(int); |\label{line:pq_decl}|

func updateEdge(src : Vertex, dst : Vertex, weight : int) |\label{line:udf_start}|
    var new_dist : int = dist[src] + weight;
    pq.updatePriorityMin(dst, dist[dst], new_dist);|\label{line:updatePriorityMin}|
end |\label{line:udf_end}|

func main()
    var start_vertex : int = atoi(argv[2]);
    dist[start_vertex] = 0;
    pq = new priority_queue |\label{line:pq_constructor_start}|
    		{Vertex}(int)(true, "lower_first", dist, start_vertex); |\label{line:pq_constructor_end}|
    while (pq.finished() == false) |\label{line:ordered_processing_operator_start}|
        var bucket : vertexset{Vertex} = pq.dequeueReadySet(); |\label{line:dequeueReadySet}|
        #s1# edges.from(bucket).applyUpdatePriority(updateEdge);|\label{line:delta_stepping_apply_update_priority}|
        delete bucket;
    end |\label{line:ordered_processing_operator_end}|
end
\end{lstlisting}
\caption{\OG algorithm for \deltastepping{} for SSSP. 
Priority-based data structures and operators are highlighted in red.}
\label{fig:code:delta_stepping_GraphIt}
\end{figure}

We propose a new priority-based extension to \OG{}
that simplifies writing parallel ordered graph algorithms.
\OG separates algorithm specifications from performance optimization
strategies. 
The user specifies the high-level algorithm with the algorithm language and uses 
a separate scheduling language to configure different performance optimizations. 
The algorithm language extension introduces a set of 
priority-based data structures and operators
to maintain execution ordering while hiding low-level details such
as synchronization, deduplication, and data structures to maintain ordering of execution. 
Figure~\ref{fig:code:delta_stepping_GraphIt} shows the implementation 
of \deltastepping{} using the priority-based extension which dequeues vertices with the lowest priority and updates their neighbors' distances in each round of the while loop. 
The while loop terminates when all the vertices' distances are finalized. 
The algorithm uses an abstract priority queue data structure, \stt{pq} (Line~\ref{line:pq_decl}), and the operators 
\stt{updatePriorityMin} (Line~\ref{line:updatePriorityMin}) and \stt{dequeueReadySet} (Line~\ref{line:dequeueReadySet}) to maintain priorities.

The priority-based extension uses a \emph{bucketing} data structure~\cite{Dhulipala:2017,DBLP:journals/corr/BeamerAP15} to maintain the execution ordering. 
Each bucket stores
active vertices of the same priority, and the buckets are sorted in
priority order. The program processes one bucket at a time in priority
order and dynamically moves active vertices to new buckets when their
priorities change.  
Updates to the bucket structure can be implemented
using either an \textit{eager bucket
update}~\cite{DBLP:journals/corr/BeamerAP15} approach or a
\textit{lazy bucket update}~\cite{Dhulipala:2017} approach.  
With eager bucket updates, buckets are immediately updated when the
priorities of active vertices change. 
Lazy bucketing buffers the updates and later performs a single bucket update per vertex.
Existing frameworks supporting ordered parallel graph algorithms only
support one of the two bucketing strategies described above.
However, using a suboptimal bucketing strategy can result in more than
\emph{10$\times$ slowdown}, as we show later. 
Eager and lazy bucketing
implementations use different data structures and parallelization schemes, making it difficult to combine both
approaches within a single framework.

With the priority-based extension, programmers can switch between lazy and eager bucket update strategies 
and combine bucketing optimizations with other optimizations using the scheduling language.  
The compiler leverages program analyses and transformations to generate efficient
code for different optimizations.  
The separation of algorithm and schedule also enables us to build an autotuner for \OG{} 
that can automatically find high-performance combinations of optimizations for a given ordered algorithm and graph.

Bucketing incurs high synchronization overheads, slowing down algorithms that
spend most of their time on bucket operations.
We introduce a new performance optimization, \emph{bucket
  fusion}, which drastically reduces synchronization overheads.  In an
ordered algorithm, a bucket can be processed in multiple rounds under
a bulk synchronous processing execution model. In every round, the
current bucket is emptied and vertices whose priority are updated to
the current bucket's priority are added to the bucket. The algorithm
moves on to the next bucket when no more vertices are added to the
current bucket. The key idea of  bucket fusion  is to
\emph{fuse consecutive rounds that process the same bucket}.
Using bucket fusion in \OG{} results 
in $1.2\times$--\maxSpeedup$\times$ speedup on road networks with large diameters over existing work.

\begin{figure}[t]
\centering
    \includegraphics [width=0.5\textwidth] {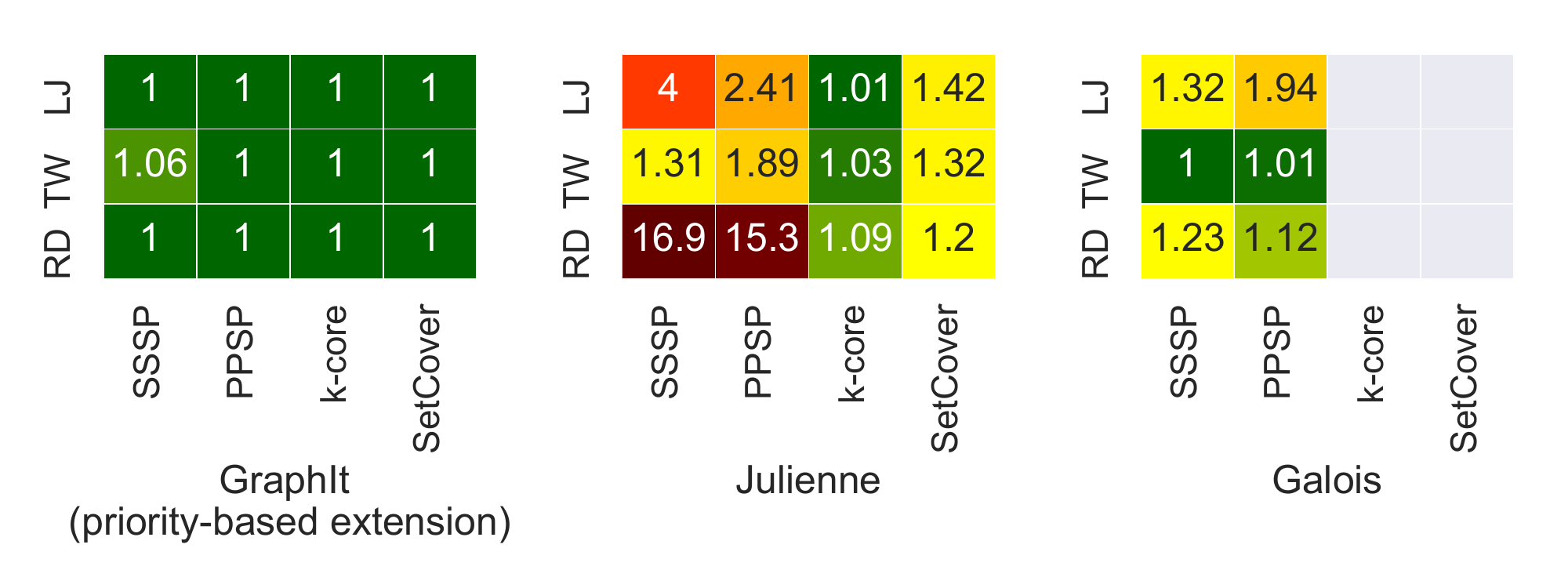}
    \caption{A heatmap of slowdowns of three frameworks compared to the fastest of all frameworks for SSSP, PPSP, \kcore, and SetCover. 
    Lower numbers (green) are better, with a value of $1$ being the fastest. Gray means that an algorithm is not supported. TW and LJ are Twitter, and LiveJournal graphs with random weights between 1 and 1000. RD is the RoadUSA graph with original weights.}
    \label{fig:heatmap}
\end{figure}

We implement the priority-based model as a language and compiler extension to \graphit~\cite{graphit:2018}\footnote{https://github.com/GraphIt-DSL/graphit}, a
domain-specific language for writing high-performance graph
algorithms. With the extension, \OG{} achieves up to \maxSpeedup$\times$ speedup on
six ordered graph algorithms (\deltastepping{} based
single-source shortest paths (SSSP), \deltastepping{} based
point-to-point shortest path (PPSP), weighted BFS (wBFS), \astar{},
\kcore{} decomposition, and approximate set cover (SetCover)) over the fastest 
state-of-the-art frameworks that support ordered algorithms
(Julienne~\cite{Dhulipala:2017} and Galois~\cite{nguyen13sosp-galois}) and hand-optimized implementations (GAPBS~\cite{DBLP:journals/corr/BeamerAP15}). 
Figure~\ref{fig:heatmap} shows that \OG is up to 16.9$\times$ and 1.94$\times$ 
faster than Julienne and Galois on the four selected algorithms and supports more algorithms than Galois.
Using \OG{}
also reduces the lines of code compared to existing frameworks and
libraries by up to 4$\times$.

The contributions of this paper are as follows.
\begin{denseitemize}
\item An analysis of the performance tradeoffs between eager and lazy bucket update optimizations
  (Sections~\ref{sec:tradeoff} and ~\ref{sec:eval}).
\item A novel performance optimization for the eager bucket update approach, \textit{bucket fusion} (Sections~\ref{sec:tradeoff} and ~\ref{sec:eval}).
\item A new priority-based programming model in \OG{} that simplifies the programming of ordered graph algorithms and makes it easy to switch between and combine different optimizations (Section~\ref{sec:programming}).
\item Compiler extensions that leverage program analyses, program transformations, and code generation to produce efficient implementations with different combinations of optimization strategies (Section~\ref{sec:compiler}).
\item A comprehensive evaluation of \OG{} that shows that it is up to \maxSpeedup$\times$ faster
  than state-of-the-art graph frameworks on six ordered graph algorithms (Section~\ref{sec:eval}). \OG{} also significantly reduces the lines of code compared to existing frameworks and libraries.
\end{denseitemize}

%% file: prelim.tex
\section{Preliminaries}

We first define \textit{ordered graph
processing} used throughout this paper.
Each vertex has a priority $p_v$. Initially, the users can explicitly
initialize priorities of vertices, with the default priority being a
null value, $\emptyset$.
These priorities change dynamically throughout the execution.
However, the priorities can only
change \emph{monotonically}, that is they can only be increased, or
only be decreased.
We say that a vertex is \emph{finalized} if its priority can no longer be
updated.
The vertices are processed and finalized based on the sorted priority ordering.  
By default, the ordered execution will stop when all vertices with
non-null priority values are finalized.
Alternatively, the user can define a customized stop condition, for
example to halt once a certain vertex has been finalized.

We define \textit{\priorityCoarsening} as an optimization to coarsen the priority value of the vertex to $p_v^\prime$ by dividing the original priority by a coarsening factor $\Delta$ such that $p_v^\prime = \lfloor p_v / \Delta \rfloor$.
The optimization is inspired by \deltastepping{} for SSSP, and enables greater parallelism at the cost of losing some algorithmic work-efficiency.
\PriorityCoarsening is used in algorithms that tolerate some priority inversions, such as \astar{}, SSSP, and PPSP, but not in \kcore{} and SetCover.

%% file: tradeoff.tex
\section{Performance Optimizations for Ordered Graph Algorithms} 
\label{sec:tradeoff}

\algrenewcommand{\alglinenumber}[1]{\footnotesize#1}
\algdef{SE}[SUBALG]{Indent}{EndIndent}{}{\algorithmicend\ }%
\algtext*{Indent} \algtext*{EndIndent}
\algblockdefx[pfor]{ParFor}{EndParFor}[1] {\textbf{parallel for}~#1~\textbf{do}}{}
\algtext*{EndParFor}

%


We use \deltastepping{} for single-source shortest
paths (SSSP) as a running example to illustrate the performance
tradeoffs between the lazy and eager bucket update approaches, and
to introduce our new bucket fusion optimization.

\begin{figure}[t]
\begin{algorithmic}[1]
\scriptsize

\State Dist $= \{\infty, \ldots, \infty\}$  \Comment{Length $|V|$ array}
\Procedure{SSSP with \deltastepping{}}{Graph $G$, $\Delta$, startV}
  \State $B$ = new LazyBucket(Dist, $\Delta$, startV);\label{line:bucket_init}
  \State Dist[startV] = 0
  \While {$\lnot$empty $B$} \label{line:outer_while_start}
	\State minBucket $= B$.getMinBucket()\label{line:min_bucket}
	 \State buffer = new BucketUpdateBuffer(); \label{line:buffer_init}
	\ParFor {src : minBucket}
		\For {e : G.getOutEdge[src]}
			\State Dist[e.dst] = min(Dist[e.dst], Dist[src] + e.weight) \label{line:update_priority}
			\State buffer.syncAppend(e.dst, $\lfloor \mbox{Dist[e.dst]}/\Delta \rfloor$)  \label{line:buffer_update}
		\EndFor
	\EndParFor
	\State buffer = buffer.reduceBucketUpdates(); \label{line:bucket_update_reduction}
	\State B.bulkUpdateBuckets(buffer);\label{line:bulk_bucket_update}
  \EndWhile \label{line:outer_while_end}
\EndProcedure
\end{algorithmic}
\caption{\deltastepping{} for single-source shortest paths (SSSP) with the lazy bucket update approach.}
\label{alg:lazy_delta_stepping}
\end{figure}

\subsection{Lazy Bucket Update}

We first consider using the lazy bucket update approach for the
\deltastepping{} algorithm, with pseudocode shown in Figure~\ref{alg:lazy_delta_stepping}.
The algorithm constructs a bucketing data structure in Line~\ref{line:bucket_init}, which groups the vertices into \emph{buckets} according to their priority.
It then repeatedly extracts the bucket with the minimum priority  (Line~\ref{line:min_bucket}), and finishes the computation once all of the buckets have been processed (Lines~\ref{line:outer_while_start}--\ref{line:outer_while_end}).
To process a bucket, the algorithm iterates over each vertex in the bucket, and updates the priority of its outgoing neighbor destination vertices by updating the neighbor's distance (Line~\ref{line:update_priority}).
With \priorityCoarsening, the algorithm computes the new priority by dividing the distance by the coarsening factor, $\Delta$.
The corresponding bucket update (the vertex and its updated priority) is added to a buffer with a synchronized append (Line~\ref{line:buffer_update}).
The \texttt{syncAppend} can be implemented using atomic operations, or with a prefix sum to avoid atomics.
The buffer is later reduced so that each vertex will only have one final bucket update (Line~\ref{line:bucket_update_reduction}).
Finally, the buckets are updated in bulk with \texttt{bulkUpdateBuckets} (Line~\ref{line:bulk_bucket_update}).




The lazy bucket update approach can be very efficient when a vertex changes buckets multiple times within a round.
The lazy approach buffers the bucket updates, and makes a single insertion to the final bucket.
Furthermore, the lazy approach can be combined with other optimizations such as histogram-based reduction on priority updates to further reduce runtime overheads.
However, the lazy approach adds additional runtime overhead from maintaining a buffer (Line~\ref{line:buffer_init}), and performing reductions on the buffer (Line~\ref{line:bucket_update_reduction}) at the end of each round.
These overheads can incur a significant cost in cases where there are only a few updates per round (e.g., in SSSP on large diameter road networks).

\subsection{Eager Bucket Update}

Another approach for implementing \deltastepping{} is to use an eager bucket update approach (shown in Figure~\ref{alg:eager_delta_stepping}) that directly updates the bucket of a vertex when its priority changes.
The algorithm is naturally implemented using thread-local buckets, which are updated in parallel across different threads (Line~\ref{line:eager_parallelization}). Each thread works on a disjoint subset of vertices in the current bucket (Line~\ref{line:eager_split}).
Using thread-local buckets avoids atomic synchronization overheads on bucket updates (Lines~\ref{line:local_bucket_init} and~\ref{line:local_bucket_update_1}--\ref{line:local_bucket_update_2}).
To extract the next bucket, the algorithm first identifies the smallest priority across all threads and then has each thread copy over its local bucket of that priority to a global minBucket (Line~\ref{line:eager_copying}).
If a thread does not have a local bucket of the next smallest priority, then it will skip the copying process.
Copying local buckets into a global bucket helps redistribute the
work among threads for better load balancing.

\begin{figure}[t]
\begin{algorithmic}[1]
\scriptsize
\State Dist $= \{\infty, \ldots, \infty\}$  \Comment{Length $|V|$ array}
\Procedure{SSSP with \deltastepping{}}{Graph $G$, $\Delta$, startV}
  \State $B$ = new ThreadLocalBuckets(Dist, $\Delta$, startV);\label{line:local_bucket_init}
  \For {threadID : threads}
    \State $B$.append(new LocalBucket());
  \EndFor

  \State Dist[startV] = 0
  \While {$\lnot$empty $B$} \label{line:outer_while}
	\State minBucket $= B$.getGlobalMinBucket() \label{line:eager_copying}
	\ParFor {threadID : threads}\label{line:eager_parallelization}
	    \For {src : minBucket.getVertices(threadID)} \label{line:eager_split}
		    \For {e : G.getOutEdge[src]}
			    \State Dist[e.dst] = min(Dist[e.dst], Dist[src] + e.weight) \label{line:local_bucket_update_1}
			    \State $B$[threadID].updateBucket(e.dst, $\lfloor \mbox{Dist[e.dst]}/\Delta \rfloor$) \label{line:local_bucket_update_2}
		    \EndFor
		\EndFor
	\EndParFor
  \EndWhile
\EndProcedure
\end{algorithmic}
\caption{\deltastepping{} for SSSP with the
  eager bucket update approach.}
\label{alg:eager_delta_stepping}
\end{figure}

Compared to the lazy bucket update approach, the eager approach saves instructions and one global synchronization needed for reducing bucket updates in the buffer (Figure~\ref{alg:lazy_delta_stepping}, Line~\ref{line:bucket_update_reduction}).
However, it potentially needs to perform multiple bucket updates per vertex in each round.

\input{opt}

%% file: opt.tex
\subsection{Eager Bucket Fusion Optimization}
\label{sec:opt}

\begin{figure}[t]
\begin{algorithmic}[1]
\scriptsize
\State Dist $= \{\infty, \ldots, \infty\}$  \Comment{Length $|V|$ array}
\Procedure{SSSP with \deltastepping{}}{Graph $G$, $\Delta$, startV}
  \State $B$ = new ThreadLocalBuckets(Dist, $\Delta$, startV);

  \For {threadID : threads}
    \State $B$.append(new LocalBucket());
  \EndFor

  \State Dist[startV] = 0
  \While {$\lnot$empty $B$}
	\State minBucket $= B$.getMinBucket()
	\ParFor {threadID : threads}
	    \For {src : minBucket.getVertices(threadID)} \label{line:bucket_processing_begin}
		    \For {e : G.getOutEdge[src]}
			    \State Dist[e.dst] = min(Dist[e.dst], Dist[src] + e.weight)
			    \State $B$[threadID].updateBucket(e.dst, $\lfloor \mbox{Dist[e.dst]}/\Delta \rfloor$) 
		    \EndFor
		\EndFor \label{line:bucket_processing_end}

	\While{$B$[threadID].currentLocalBucket() is not empty } \label{line:bucket_merge_while}
	\State currentLocalBucket = $B$[threadID].currentLocalBucket()
	\If{currentLocalBucket.size() $<$ threshold}\label{line:threshold_check}
	    \For {src : currentLocalBucket} \label{line:fused_bucket_processing_begin}
		    \For {e : G.getOutEdge[src]}
			    \State Dist[e.dst] = min(Dist[e.dst], Dist[src] + e.weight)
			    \State $B$[threadID].updateBucket(e.dst, $\lfloor \mbox{Dist[e.dst]}/\Delta \rfloor$) 
		    \EndFor
		\EndFor \label{line:fused_bucket_processing_end}

	\Else
	    \ break
	\EndIf
	\EndWhile

	\EndParFor
  \EndWhile
\EndProcedure
\end{algorithmic}
\caption{\deltastepping{} for single-source shortest paths with the eager bucket update approach and the bucket fusion optimization.}
\label{alg:eager_delta_stepping_merge}
\end{figure}


A major challenge in bucketing is that a large number of buckets need
to be processed, resulting in thousands or even tens of thousands of
processing rounds.  Since each round requires at
least one global synchronization, reducing the number of rounds while
maintaining priority ordering can significantly reduce synchronization
overhead.

Often in practice, many consecutive rounds process
a bucket of the same priority.
For example, in \deltastepping{}, the priorities of vertices that are
higher than the current priority can be lowered by edge relaxations to
the current priority in a single round. As a result, the same
priority bucket may be processed again in the next round.  The process
repeats until no new vertices are added to the current bucket.
This pattern is common in ordered graph algorithms that use
\priorityCoarsening.
We observe that rounds processing the same
bucket can be fused without violating priority ordering.


Based on this observation, we propose a novel bucket fusion
optimization for the eager bucket update approach that allows a thread
to execute the next round processing the current bucket without synchronizing with other
threads.
We illustrate bucket fusion using the \deltastepping{} algorithm in Figure~\ref{alg:eager_delta_stepping_merge}.
The same optimization can be applied in other applications, such as wBFS, \astar{} and point-to-point shortest path.
This algorithm extends the eager bucket update algorithm
(Figure~\ref{alg:eager_delta_stepping}) by adding a while loop inside each
local thread (Figure~\ref{alg:eager_delta_stepping_merge}, Line~\ref{line:bucket_merge_while}). 
The while loop executes if the current local bucket is non-empty. 
If the current local bucket's size is below a certain
threshold, the algorithm immediately processes the current bucket without synchronizing with other threads
(Figure~\ref{alg:eager_delta_stepping_merge}, Line~\ref{line:threshold_check}).  
If the current local bucket is large, it will be copied over to the global bucket and
distributed across other threads. 
The threshold is important to avoid
creating straggler threads that process too many vertices, leading to
load imbalance.  The bucket processing logic in the while loop
(Figure~\ref{alg:eager_delta_stepping_merge}, Lines~\ref{line:fused_bucket_processing_begin}--\ref{line:fused_bucket_processing_end})
is the same as the original processing logic (Figure~\ref{alg:eager_delta_stepping_merge}, 
Lines~\ref{line:bucket_processing_begin}--\ref{line:bucket_processing_end}).
This optimization is hard to apply for the lazy approach since a
global synchronization is needed before bucket updates.


Bucket fusion is particularly useful for road networks where multiple rounds frequently process the same bucket.
For example, bucket fusion reduces the number of rounds by more than 30$\times$ for SSSP on the RoadUSA graph, leading to more than 3$\times$ speedup by significantly reducing the amount of global synchronization (details in Section~\ref{sec:eval}).


%% file: programming.tex
\section{Programming Model}
\label{sec:programming}

\begin{table*}[t]
\centering
 \footnotesize
 \caption{Algorithm language extensions in \OG{}.}
 \begin{tabular}{p{6.5cm}p{1.5cm}p{8.5cm}}
 \textbf{Edgeset Apply Operator}  & \textbf{Return Type} & \textbf{Description} \\ 
\hline

\ftt{applyUpdatePriority(func f)} & none & Applies \texttt{f}(src, dst) to every edge. The \texttt{f} function updates priorities of vertices. \\
 \medskip
  \textbf{Priority Queue Operators}  & &  \\ 
\hline
    \ftt{new priority\_queue( \newline \hspace*{0.5ex} bool allow\_\priorityCoarsening,
    \newline \hfill \hspace*{0.5ex} string priority\_direction,
    \newline \hspace*{0.5ex} vector priority\_vector,
    \newline \hspace*{0.5ex} Vertex optional\_start\_vertex)} & priority\_queue &  The constructor for the priority queue. It specifies whether \priorityCoarsening is allowed or not, higher or lower priority gets executed first, the vector that is used to compute the priority values, and an optional start vertex. A \texttt{lower\_first} priority direction specifies that lower priority values are executed first, whereas a \texttt{higher\_first} indicates higher priority values are executed first.\\ \hline
   \ftt{pq.dequeueReadySet()} & \textsf{vertexset} & Returns a bucket with all the vertices that are currently ready to be processed.  \\ \hline
   \ftt{pq.finished()} &  bool & Checks if there is any bucket left to process.\\ \hline
   \ftt{pq.finishedVertex(Vertex v)} & bool & Checks if a vertex's priority is finalized (finished processing).\\ \hline
   \ftt{pq.getCurrentPriority()} &  priority\_type & Returns the priority of the current bucket.\\ \hline
   \ftt{pq.updatePriorityMin(Vertex v, ValT new\_val)} & void & Decreases the value of the priority of the specified vertex \textsf{v} to the new\_val.\\ \hline
    \ftt{pq.updatePriorityMax(Vertex v, ValT new\_val)} & void & Increases the value of the priority of the specified vertex \textsf{v} to the new\_val.\\ \hline
   \ftt{pq.updatePrioritySum(Vertex v, ValT sum\_diff, \hspace*{0.5ex} ValType min\_threshold)} & void & Adds sum\_diff to the priority of the Vertex \textsf{v}. The user can specify an optional minimum threshold so that the priority will not go below the threshold.  \\ \hline

\noalign{\smallskip}
  \end{tabular}
  \label{table:algo_api}
\end{table*}

The \extension{} follows the design of \OG{} and separates the algorithm specification from the performance optimizations, similar to Halide~\cite{Ragan-Kelley:2013:HLC:2499370.2462176}
and Tiramisu~\cite{Baghdadi:2019:TPC}.
The user writes a high-level algorithm using the algorithm language and specifies optimization strategies
 using the scheduling language.
The extension introduces a set of
priority-based data structures and operators to \OG{}
to maintain execution order in the algorithm language and adds support for bucketing optimizations in the scheduling language.

\subsection{Algorithm Language}

The algorithm language exposes opportunities for eager bucket update, eager update with bucket fusion, lazy bucket update, and other optimizations.
The high-level operators hide low-level implementation details such as atomic synchronization, deduplication, bucket updates, and \priorityCoarsening.
The algorithm language shares the vertex and edge sets, and operators that apply user-defined functions on the sets with the \graphit algorithm language.

The \extension{} proposes high-level priority queue-based abstractions to switch between thread-local and global buckets. The extension to \OG{} also introduces priority update operators to hide the bucket update mechanisms, and provides a new \textsf{edgeset} apply operator, \stt{applyUpdatePriority}. The priority-based data structures and operators are shown in Table~\ref{table:algo_api}.

Figure~\ref{fig:code:delta_stepping_GraphIt} shows an example of \deltastepping{} for SSSP.
\OG works on vertex and edge sets. The algorithm specification first sets up the \textsf{edgeset} data structures (Lines~\ref{line:vertexset_edgeset_setup_begin}--\ref{line:vertexset_edgeset_setup_end}), and sets the distances to all the vertices in \stt{dist} to INT\_MAX to represent $\infty$ (Line~\ref{line:dist_init}).
It declares the global priority queue, \stt{pq}, on Line~\ref{line:pq_decl}.
This priority queue can be referenced in user-defined functions and the main function.
The user then defines a function, \stt{updateEdge}, that processes each edge (Lines~\ref{line:udf_start}--\ref{line:udf_end}).
In \stt{updateEdge}, the user computes a new distance value, and then updates the priority of the destination vertex using the \stt{updatePriorityMin} operator defined in Table~\ref{table:algo_api}.
In other algorithms, such as $k$-core,
the user can use \stt{updatePrioritySum} when the priority is decremented or incremented by a given value.
The \stt{updatePrioritySum} can detect if the change to the priority is a constant, and use this fact to perform more optimizations.
The priority update operators, \stt{updatePriorityMin} and \stt{updatePrioritySum}, hide bucket update operations, allowing the compiler to generate different code
for lazy and eager bucket update strategies.



Programmers use the constructor of the priority queue (Lines~\ref{line:pq_constructor_start}--\ref{line:pq_constructor_end}) to specify algorithmic information, such as the priority ordering, support for priority coarsening, and the direction that priorities change (documented in Table~\ref{table:algo_api}).
The abstract priority queue hides low-level bucket implementation details and provides a mapping between vertex data and their priorities.
The user specifies a \stt{priority\_vector} that stores the vertex data values used for computing priorities. In SSSP, we use the \stt{dist} vector and the coarsening parameter ($\Delta$ specified using the scheduling language) to perform \priorityCoarsening.
The while loop (Line~\ref{line:ordered_processing_operator_start}) processes vertices from a bucket until all buckets are finished processing.
In each iteration of the while loop, a new bucket is extracted with \stt{dequeueReadySet} (Line~\ref{line:dequeueReadySet}).
The \textsf{edgeset} operator on Line~\ref{line:delta_stepping_apply_update_priority} uses the \stt{from} operator to keep only the edges that come out of the vertices in the bucket. Then it uses \stt{applyUpdatePriority}
to apply the \stt{updateEdge} function to outgoing edges of the bucket. Label ({\small \textbf{\stt{\#s1\#}}}) is later used by the scheduling language to configure optimization strategies.


\begin{figure} [t]
\begin{lstlisting} [firstnumber=17,language=graphit,escapechar=|]
	...
    while (pq.finished() == false)
        var bucket : vertexsubset = pq.dequeueReadySet();
        #s1# edges.from(bucket).applyUpdatePriority(updateEdge);
        delete bucket;
    end |\Suppressnumber|
...|\Reactivatenumber{25}|
schedule:
program->configApplyPriorityUpdate("s1", "lazy")
->configApplyPriorityUpdateDelta("s1", "4")
->configApplyDirection("s1", "SparsePush")
->configApplyParallelization("s1","dynamic-vertex-parallel");
\end{lstlisting}
\caption{\OG{} scheduling specification for \deltastepping{}.}
\label{lst:delta_stepping_schedule}
\end{figure}

\subsection{Scheduling Language}

\begin{table*}[t]
\centering
 \footnotesize
  \caption{Scheduling functions for \stt{applyUpdatePriority} operators. Default options are in bold. }
 \begin{tabular}{p{6.3cm}p{10.2cm}}
\textbf{Apply Scheduling Functions}  & \textbf{Descriptions} \\ \hline
\ftt{configApplyPriorityUpdate(label,config);} & Config options: eager\_with\_fusion, eager\_no\_fusion, lazy\_constant\_sum, and \textbf{lazy}. \\ \hline
\ftt{configApplyPriorityUpdateDelta(label,config);} & Configures the $\Delta$ parameter for coarsening the priority range. \\ \hline
\ftt{configBucketFusionThreshold(label, config);} & Configures the threshold for the bucket fusion optimization.\\ \hline
\ftt{configNumBuckets(label,config);} & Configures the number of buckets that are materialized for the lazy bucket update approach.\\ \hline
  \end{tabular}
  \label{table:schedule_api}
\end{table*}

\begin{figure*}[t]
\centering
    \includegraphics [width=\textwidth] {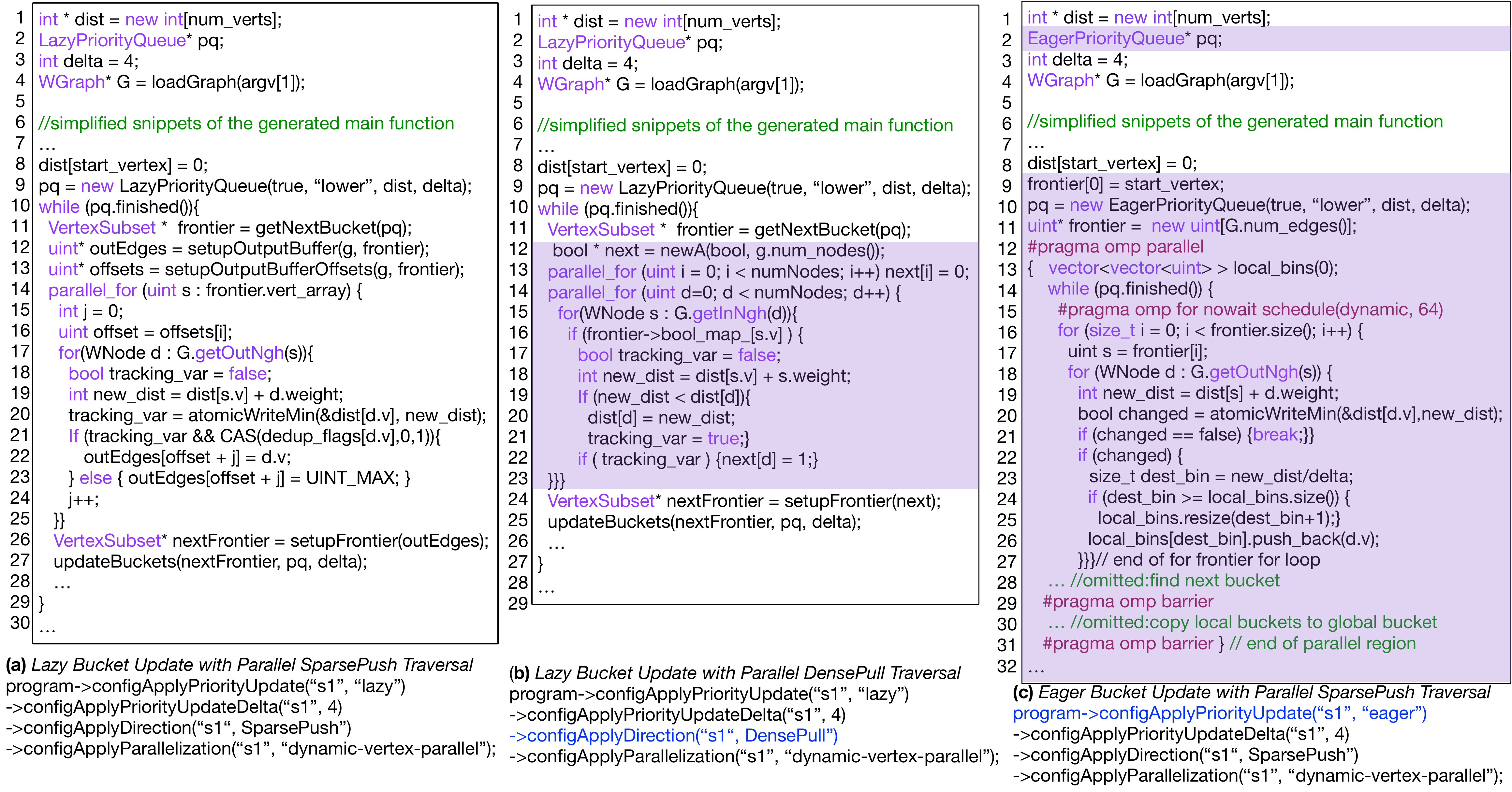}
    \caption{Simplified generated C++ code for \deltastepping{} for SSSP with different schedules. Changes in schedules for (b) and (c) compared to (a) are highlighted in blue. Changes in the generated code are highlighted in purple background. \texttt{parallel\_for} is translated to \stt{cilk\_for} or \stt{\#pragma omp parallel for}. Struct \stt{WNode} has two fields, \stt{v} and \stt{weight}. \stt{v} stores the vertex ID and \stt{weight} stores the edge weight.}
    \label{fig:sssp_different_schedules}
\end{figure*}


The scheduling language allows users to specify different
optimization strategies without changing the algorithm.
We extend the scheduling language of \graphit with new commands to enable switching between eager and lazy bucket update strategies.
Users can also tune other parameters, such as the coarsening factor for \priorityCoarsening.
The scheduling API extensions are shown in Table~\ref{table:schedule_api}.

Figure~\ref{lst:delta_stepping_schedule} shows a set of schedules for \deltastepping{}.
\OG uses labels ({\small \textbf{\stt{\#label\#}}}) to identify the algorithm language statements for which the scheduling language commands are applied.
The programmer adds the label \textbf{s1} to the \textsf{edgeset} \stt{applyUpdatePriority} statement.
After the \scheduleKeyword{} keyword, the programmer calls the
scheduling functions. The
\stt{configApplyPriorityUpdate} function allows the programmer to use the lazy bucket update optimization.
The programmer can use the original \graphit scheduling language to configure the direction of edge traversal (\stt{configApplyDirection}) and the load balance strategies (\stt{configApplyParallelization}).
Direction optimizations can be combined with lazy priority update schedules.
\stt{configApplyUpdateDelta} is used to set the delta for \priorityCoarsening.

Users can change the schedules to generate code with different combinations of optimizations as shown in Figure~\ref{fig:sssp_different_schedules}.
Figure~\ref{fig:sssp_different_schedules}(a) shows code generated by combining the lazy bucket update strategy and other edge traversal optimizations from the \graphit scheduling language.
The scheduling function \stt{configApplyDirection} configures the data layout of the frontier and direction of the edge traversal (\stt{SparsePush} means sparse frontier and push direction).
Figure~\ref{fig:sssp_different_schedules}(b) shows the code generated when we combine a different traversal direction (\stt{DensePull}) with the lazy bucketing strategy.
Figure~\ref{fig:sssp_different_schedules}(c) shows code generated with the eager bucket update strategy. Code generation is explained in Section~\ref{sec:compiler}.

%% file: compiler.tex
\section{Compiler Implementation}
\label{sec:compiler}

We demonstrate how the compiler generates code for different bucketing optimizations.
The key challenges are in how to insert low-level synchronization and deduplication instructions, and how to combine bucket optimizations with direction optimization and other optimizations in the original \graphit scheduling language.
Furthermore, the compiler has to perform global program transformations and code generation to switch between lazy and eager approaches.


\subsection{Lazy Bucket Update Schedules}
To support the lazy bucket update approach, the compiler applies program analyses and transformations on the user-defined functions (UDFs).
The compiler uses dependence analysis on \stt{updatePriorityMin} and \stt{updatePrioritySum} to determine if there are write-write conflicts and insert atomics instructions as necessary (Figure~\ref{fig:sssp_different_schedules}(a) Line 20).
Additionally, the compiler needs to insert variables to track whether a vertex's priority has been updated or not (\stt{tracking\_var} in Figure~\ref{fig:sssp_different_schedules}(a), Line 18).
This variable is used in the generated code to determine which
vertices should be added to the buffer \stt{outEdges} (Figure~\ref{fig:sssp_different_schedules}(a), Line 21).
Deduplication is enabled with a compare-and-swap (CAS) on deduplication flags (Line 21) to ensure that each vertex is inserted into the outEdges only once.
Deduplication is required for correctness for applications such as \kcore{}.
Changing the schedules with different traversal directions or frontier layout affects the code generation for edge traversal and user-defined functions (Figure~\ref{fig:sssp_different_schedules}(b)).
In the \DensePull traversal direction, no atomics are needed for the destination nodes.

We built runtime libraries to manage the buffer and update buckets. The compiler
generates appropriate calls to the library (\stt{getNextBucket}, \stt{setupFrontier}, and \stt{updateBuckets}).
The \stt{setupFrontier} API (Figure~\ref{fig:sssp_different_schedules}(a), Line 24) performs a prefix sum on the \stt{outEdges} buffer to compute the next frontier.
We use a lazy priority queue (declared in Figure~\ref{fig:sssp_different_schedules}(a), Line 2) for storing active vertices based on their priorities.
The lazy bucketing is based on Julienne's bucket data structures that only materialize a few buckets, and keep vertices outside of the current range in an overflow bucket~\cite{Dhulipala:2017}.
We improve its performance by redesigning the lazy priority queue interface.
Julienne's original interface invokes a lambda function call to compute the priority.
The \extension{} computes the priorities using a priority vector and $\Delta$ value for priority coarsening, eliminating extra function calls.

\begin{figure}[t]
\begin{lstlisting} [language=graphit,escapechar=|]
func apply_f(src: Vertex, dst: Vertex)
    var k: int = pq.get_current_priority();
    pq.updatePrioritySum(dst, -1, k); |\label{line:constant_update} |
end
\end{lstlisting}

    \begin{lstlisting} [language=c++,escapechar=|]
apply_f_transformed = [&] (uint vertex, uint count) { | \label{line:count_arg} |
    int k = pq->get_current_priority();
    int priority = pq->priority_vector[vertex];
    if (priority > k) {
      uint __new_pri = std::max(priority + (-1) * count, k);
      pq->priority_vector[vertex] = __new_pri;
      return wrap(vertex, pq->get_bucket(__new_pri));}}
    \end{lstlisting}
    \caption{The original (top) and transformed (bottom) user-defined function for \kcore{} using lazy with constant sum reduction.}
    \label{fig:code:kcore_udf}
\end{figure}

\myparagraph{Lazy with constant sum reduction} We also incorporated a specialized histogram-based reduction optimization (first proposed in Julienne~\cite{Dhulipala:2017}) to reduce priority updates with a constant value each time.
This optimization can be combined with the lazy bucket update strategy to improve performance.
For \kcore{}, since the priorities for each vertex always reduce by one at each update, we can optimize it further by keeping track of only the number of updates with a histogram.
This way, we avoid contention on vertices that have a large number of neighbors on the frontier.

To generate code for the histogram optimization, the compiler first analyzes the user-defined function to determine whether the change to the priority of the vertex is a fixed value and if it is a sum reduction (Figure~\ref{fig:code:kcore_udf} (top), Line~\ref{line:constant_update}).
The compiler ensures that there is only one priority update operator in the user-defined function.
It then extracts the fixed value (\stt{-1}), the minimum priority (\stt{k}), and vertex identifier (\stt{dst}).
In the transformed function (Figure~\ref{fig:code:kcore_udf} (bottom)), an if statement and max operator are generated to check and maintain the priority of the vertex.
The \stt{applyUpdatePriority} operator gets the counts of updates to each vertex using a histogram approach and supplies the vertex and count as arguments to the transformed function (Figure~\ref{fig:code:kcore_udf} (bottom), Line~\ref{line:count_arg}).
The compiler copies all of the expressions used in the priority update operator and the expressions that they depend on in the transformed function.




\subsection{Eager Bucket Update Schedules}
The compiler uses program analysis to determine feasibility of the transformation, transforms user-defined functions and edge traversal code, and uses optimized runtime libraries to generate efficient code for the eager bucket update approach.

The compiler analyzes the while loop (Figure~\ref{fig:code:delta_stepping_GraphIt}, Lines~\ref{line:ordered_processing_operator_start}--\ref{line:ordered_processing_operator_end}) to look for the pattern of an iterative priority update with a termination criterion.
The analysis checks that there is no other use of the generated vertexset (\texttt{bucket}) except for the \texttt{applyUpdatePriority} operator, ensuring correctness.

Once the analysis identifies the while loop and edge traversal operator, the compiler replaces the while loop with an ordered processing operator.
The ordered processing operator uses an OpenMP parallel region (Figure~\ref{fig:sssp_different_schedules}(c), Lines 12--32) to set up thread-local data structures, such as \stt{local\_bins}.
We built an optimized runtime library for the ordered processing operator based on the \deltastepping{} implementation in GAPBS~\cite{DBLP:journals/corr/BeamerAP15}.
A global vertex frontier (Figure~\ref{fig:sssp_different_schedules}(c), Line 11) keeps track of vertices of the next priority (the next bucket).
In each iteration of the while loop, the \texttt{\#pragma omp for} (Figure~\ref{fig:sssp_different_schedules}(c), Lines 15--16) distributes work among the threads. After priorities and buckets are updated, each local thread proposes its next bucket priority, and the smallest priority across threads will be selected (omitted code on Figure~\ref{fig:sssp_different_schedules}(c), Line 28).
Once the next bucket priority is decided, each thread will copy vertices in its next local bucket to the global frontier (Figure~\ref{fig:sssp_different_schedules}(c), Line 30)

Finally, the compiler transforms the user-defined functions by appending the local buckets to the argument list and inserting appropriate synchronization instructions. These transformations allow priority update operators to directly update thread-local buckets (Figure~\ref{fig:sssp_different_schedules}(c), Lines 23--26).

\myparagraph{Bucket Fusion}
The bucket fusion optimization adds another while loop after end of the for-loop on Line 27 of Figure~\ref{fig:sssp_different_schedules}(c), and before finding the next bucket across threads on Line 28.
This inner while loop processes the current bucket in the local priority queue (\stt{local\_bins}) if it is not empty and its size is less than a threshold.
In the inner while loop, vertices are processed using the same transformed user-defined functions as before.
The size threshold improves load balancing, as large buckets are distributed across different threads.

\subsection{Autotuning}
We built an autotuner on top of the extension to automatically find high-performance schedules for a given algorithm and graph.
The autotuner is built using OpenTuner~\cite{ansel:pact:2014} and stochastically searches through a large number of optimization strategies generated with the scheduling language.
It uses an ensemble of search methods, such as the area under curve bandit meta technique, to find good combinations of optimizations within a reasonable amount of time.

%% file: eval.tex
\section{Evaluation}
\label{sec:eval}
We compare the performance of the \extension in \OG{} to state-of-the-art
frameworks and analyze performance tradeoffs among different \OG schedules.
We use a dual-socket system with Intel Xeon E5-2695 v3 CPUs with 12
cores each for a total of 24 cores and 48 hyper-threads. The system
has 128 GB of DDR3-1600 memory and 30 MB last level cache on each
socket and runs with Transparent Huge Pages (THP) enabled and Ubuntu 18.04.

\begin{table}[t]
\begin{center}
  \scriptsize
    \caption{Graphs used for experiments. The number of edges are directed edges. Graphs are symmetrized for \kcore{} and SetCover.} 
  \begin{tabular}{p{0.9cm}| l|p{0.8cm}|p{0.8cm}|p{1cm}}
Type & Dataset & Num. Verts &  Num. Edges & Symmetric Num.Edges \\ \hline
Social  & \textit{Orkut} (OK)~\cite{friendster} & 3 M & 234 M & 234 M \\ 
 Graphs   & \textit{LiveJournal} (LJ)~\cite{davis11acm-florida-sparse} & 5 M & 69 M  & 85M \\ 
    & \textit{Twitter} (TW)~\cite{kwak10www-twitter} & 41 M & 1469 M & 2405 M\\ 
    & \textit{Friendster} (FT)~\cite{friendster} & 125 M & 3612 M & 3612 M\\ \hline
    
Web Graph    & \textit{WebGraph} (WB)~\cite{sd-graph} & 101 M & 2043 M & 3880 M\\ \hline
Road    & \textit{Massachusetts} (MA)~\cite{openstreetmap} & 0.45 M & 1.2 M & 1.2 M \\
 Graphs   & \textit{Germany} (GE)~\cite{openstreetmap} & 12 M & 32 M & 32 M \\ 
    & \textit{RoadUSA} (RD)~\cite{road-graph} & 24 M & 58 M & 58 M \\ \hline
  \end{tabular}
 \label{table:datasets}
\end{center}
\end{table}

\input{eval-perf-table}



\myparagraph{Data Sets}
Table~\ref{table:datasets} shows our input graphs and their sizes. 
For \kcore{} and SetCover, we symmetrize the input graphs.
For \deltastepping{} based SSSP, wBFS, PPSP using \deltastepping{}, and \astar{}, we use the original directed versions of graphs with integer edge weights. 
The RoadUSA (RD), Germany(GE) and Massachusetts (MA) road graphs are used for the \astar{} algorithm, as they have the longitude and latitude data for each vertex. 
GE and MA are constructed from data downloaded from OpenStreetMap~\cite{openstreetmap}. 
Weight distributions used for experiments are described in the caption of Table~\ref{table:eval-perf-table}.


\myparagraph{Existing Frameworks} 
Galois v4~\cite{nguyen13sosp-galois} uses approximate priority ordering with an ordered list abstraction for SSSP. 
We implemented PPSP and \astar{} using the ordered list. 
To the best of our knowledge and from communications with the developers, strict priority-based ordering is currently not supported for Galois. 
Galois does not provide implementations of wBFS, \kcore{} and SetCover, which require strict priority ordering.
GAPBS~\cite{DBLP:journals/corr/BeamerAP15} is a suite of C++ implementations of graph algorithms and uses eager bucket update for SSSP.
GAPBS does not provide implementations of \kcore{} and SetCover.
We used Julienne~\cite{Dhulipala:2017} from early 2019.
The developers of Julienne have since 
incorporated the optimized bucketing interface proposed in this paper in the latest version. 
\graphit~\cite{graphit:2018} and Ligra~\cite{shun13ppopp-ligra} are two of the fastest unordered graph frameworks. 
We used the best configurations (e.g., priority coarsening factor $\Delta$ and the number of cores) for the comparison frameworks. Schedules and parameters used are in the artifact.

\subsection{Applications}

We evaluate the extension to \OG on SSSP with \deltastepping{}, weighted breadth-first search (wBFS), point-to-point shortest path (PPSP), \astar{}, $k$-core decomposition (\kcore{}), and approximate set cover (SetCover).

\myparagraph{SSSP and Weighted Breadth-First Search (wBFS)} SSSP with
\deltastepping{} solves the single-source shortest path problem as shown in Figure~\ref{alg:lazy_delta_stepping}. In \deltastepping{}, vertices are partitioned into buckets with interval $\Delta$ according to their current shortest distance. In each iteration, the smallest non-empty bucket $i$ which contains all vertices with distance in $[{i\Delta, (i+1)\Delta})$ is processed. 
 wBFS is a special case of \deltastepping{} for graphs with positive integer edge weights, with delta fixed to 1~\cite{Dhulipala:2017}. 
We benchmarked wBFS on only the social networks and web graphs with weights in the range $[1,\log n)$, following the convention in previous work~\cite{Dhulipala:2017}.


\myparagraph{Point-to-point Shortest Path (PPSP)}
Point-to-point shortest path (PPSP) takes a graph $G(V, E, w(E))$, a source vertex $s\in V$, and a destination vertex $d\in V$ as inputs and computes the shortest path between $s$ and $d$. In our PPSP application, we used the \deltastepping{} algorithm with \priorityCoarsening. It terminates the program early when it enters iteration $i$ where $i\Delta$ is greater than or equal to the shortest distance between $s$ and $d$ it has already found.

\myparagraph{A$^*$ Search}
The \astar{} algorithm finds the shortest path between two points.  
The difference between \astar{} and \deltastepping{} is that, instead of using the current shortest distance to a vertex as priority, \astar{} uses the \textit{estimated distance} of the shortest path that goes from the source to the target vertex that passes through the current vertex as the priority.
Our \astar{} implementation is based on a related work~\cite{chronos} and needs the longitude and latitude of the vertices. 



\myparagraph{\kcore{}}
A \kcore{} of an undirected graph $G(V, E)$ refers to a maximal connected sub-graph of G where all vertices in the sub-graph have induced-degree at least $k$. The \kcore{} problem takes an undirected graph $G(V, E)$ as input and for each $u\in V$ computes the \emph{maximum} \kcore{} that $u$ is contained in (this value is referred to as the coreness of the vertex) using a peeling procedure~\cite{Matula:1983:SOC:2402.322385}.


\myparagraph{Approximate Set Cover}
The set cover problem takes as input a universe $\mathcal{U}$ containing a set of ground elements, a collection of sets $\mathcal{F}$ s.t. $\cup_{f \in \mathcal{F}} f = \mathcal{U}$, and a cost function $c : \mathcal{F} \rightarrow \mathbb{R}_{+}$. The problem is to find the cheapest collection of sets $\mathcal{A} \subseteq \mathcal{F}$ that covers $\mathcal{U}$, i.e. $\cup_{a \in \mathcal{A}} a = \mathcal{U}$. In this paper, we implement the unweighted version of the problem, where $c : \mathcal{F} \rightarrow 1$, but the algorithm used easily generalizes to the weighted case~\cite{Dhulipala:2017}. The algorithm at a high-level works by bucketing the sets based on their cost per element, i.e., the ratio of the number of uncovered elements they cover to their cost. At each step, a nearly-independent subset of sets from the highest bucket (sets with the best cost per element) are chosen, removed, and the remaining sets are reinserted into a bucket corresponding to their new cost per element. We refer to the following papers by Blelloch et al.~\cite{blelloch11manis, blelloch12setcover} for algorithmic details and a survey of related work.

\subsection{Comparisons with other Frameworks}
Table~\ref{table:eval-perf-table} shows the execution times of \OG with the \extension{} and other frameworks.
\OG outperforms the next fastest of Julienne, Galois, GAPBS, \graphit, and Ligra by up to \maxSpeedup{}$\times$ and is no more than \maxSlowdown{} 
slower than the fastest. 
\OG is up to 16.8$\times$ faster than Julienne, 7.8$\times$ faster than Galois, and 
3.5$\times$ faster than hand-optimized GAPBS.
Compared to unordered frameworks, \graphit without the priority-based extension (unordered) and Ligra, \OG with the extension achieves speedups between 1.67$\times$ to more than 600$\times$ due to improved algorithm efficiency.
The times for SSSP and wBFS are averaged over 10 starting vertices. 
The times for PPSP and \astar{} are averaged over 10 source-destination pairs. 
We chose start and end points to have a balanced selection of different distances.

\OG with the priority extension has the fastest SSSP performance on six out of the seven input graphs.
Julienne uses significantly more instruction than \OG (up to 16.4$\times$ instructions than \OG).
On every iteration, Julienne computes an out-degree sum for the vertices on the frontier to use the direction optimization, which adds significant runtime overhead.
\OG{} avoids this overhead by disabling the direction optimization with the scheduling language. 
Julienne also uses lazy bucket update that generates extra instructions to buffer the bucket updates whereas \OG saves instructions by using eager bucket update.
\OG is faster than GAPBS because of the bucket fusion optimization that allows \OG to process more vertices in each round and use fewer rounds (details are shown in Table~\ref{table:fusion_perf_impact}). 
The optimization is especially effective for road networks, where synchronization overhead is a significant performance bottleneck. 
Galois achieves good performances on SSSP because it does not have as much overhead from global synchronization needed to enforce strict priority. 
However, it is slower than \OG on most graphs because approximate priority ordering sacrifices some work-efficiency.

\OG with the priority extension is the fastest on most of the graphs for PPSP, wBFS, and \astar{}, which use a variant of the \deltastepping{} algorithm with \priorityCoarsening. 
Both \OG and GAPBS use eager bucket update for these algorithms. 
\OG outperforms GAPBS because of bucket fusion. 
Galois is often slower than \OG due to lower work-efficiency with the approximate priority ordering. 
Julienne uses lazy bucket update and is slower than \OG due to the runtime overheads of the lazy approach.

PPSP and \astar{} are faster than SSSP as they only run until the distance to the destination vertex is finalized. 
\astar{} is sometimes slower than PPSP because of additional random memory accesses and computation needed for estimating distances to the destination.

For \kcore{} and SetCover, the extended \OG runs faster than Julienne because the optimized lazy bucketing interface uses the priority vector to compute the priorities of each vertex. 
Julienne uses a user-defined function to compute the priority every time, resulting in function call overheads and redundant computations. 
Galois does not provide ordered algorithms for \kcore{} and SetCover, which require strict priority and synchronizations after processing each priority. 


\myparagraph{Delta Selection for Priority Coarsening} 
The best $\Delta$ value for each algorithm depends on the size and the structure of the graph. 
The best $\Delta$ values for social networks (ranging from 1 to 100) are much smaller than deltas for road networks with large diameters (ranging from $2^{13}$ to $2^{17}$). 
Social networks need only a small $\Delta$ value because they have ample parallelism with large frontiers and work-efficiency is more important.
Road networks need larger $\Delta$ values for more parallelism. 
We also tuned the $\Delta$ values for the comparison frameworks to provide the best performance. 

\myparagraph{Autotuning} The autotuner for \OG{} is able to automatically find schedules that performed within 5$\%$ of the hand-tuned schedules used for Table~\ref{table:eval-perf-table}. 
For most graphs, the autotuner can find a high-performance schedule within 300s after trying 30-40 schedules (including tuning integer parameters) in a large space of about $10^6$ schedules. 
The autotuning process finished within 5000 seconds for the largest graphs.
Users can specify a time limit to reduce autotuning time.

\begin{table}[t]
\begin{center}
\footnotesize
    \caption{Line counts of SSSP, PPSP, \astar{}, \kcore{}, and SetCover for \OG, GAPBS, Galois, and Julienne. The missing numbers correspond to a framework not providing an algorithm.}
  \begin{tabular}{l|p{1.5cm}|c|c|c}
     & \OG with extension & GAPBS & Galois & Julienne \\ \hline
    SSSP & \textbf{28} & 77 & 90 & 65 \\ 
    PPSP & \textbf{24} & 80 & 99 & 103 \\
    A* & \textbf{74} & 105 & 139 & 84 \\ 
    KCore & \textbf{24} & --  & --  & 35 \\
    SetCover & \textbf{70} & --  & -- & 72 \\ \hline
  \end{tabular}
  \label{table:line-of-code}
\end{center}
\end{table}

\myparagraph{Line Count Comparisons} Table~\ref{table:line-of-code} shows the line counts of the five graph algorithms implemented in four frameworks. 
GAPBS, Galois, and Julienne all require the programmer to take care of implementation details such as atomic synchronization and deduplication. \OG uses the compiler to automatically generate these instructions.
For \astar{} and SetCover, \OG needs to use long \texttt{extern} functions that significantly increases the line counts.

\subsection{Scalability Analysis}
\begin{figure}[t]
\centering
    \includegraphics [width=0.47\textwidth] {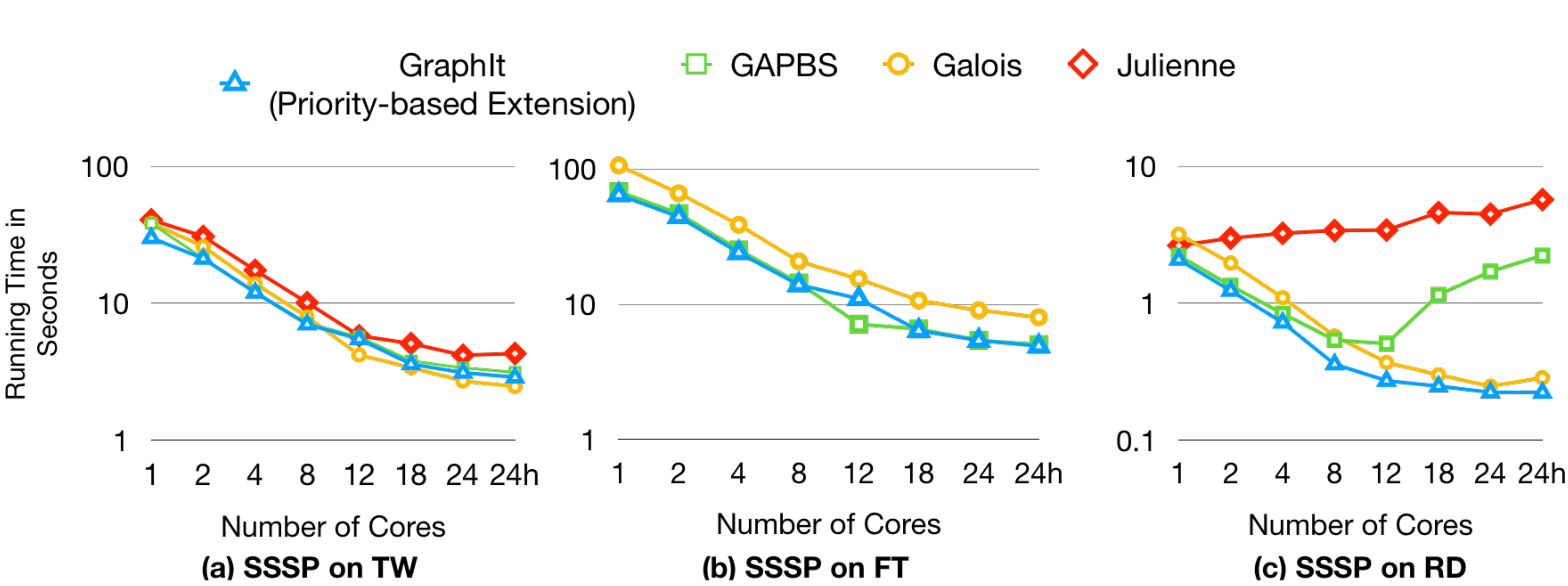}
    \caption{Scalability of different frameworks on SSSP.}
    \label{fig:scalability}
\end{figure}
We analyze the scalability of different frameworks in Figure~\ref{fig:scalability} for SSSP on social and road networks. 
The social networks (TW and FT) have very small diameters 
and large numbers of vertices. 
As a result, they have a lot of parallelism in each bucket, 
and all three frameworks scale reasonably well (Figure~\ref{fig:scalability}(a) and (b)). 
Compared to GAPBS, \OG uses bucket fusion to significantly reduce synchronization overheads and improves parallelism on the RoadUSA network (Figure~\ref{fig:scalability}(c)). 
GAPBS suffers from NUMA accesses when going beyond a single socket (12 cores). 
Julienne's overheads from lazy bucket updates make it hard to scale on the RoadUSA graph. 

\subsection{Performance of Different Schedules}
\label{sec:schedules_perf}

\begin{table}[t]
  \center \footnotesize
  \caption{Running times and number of rounds reductions with the bucket fusion optimization on SSSP using \deltastepping{}.}
  \begin{tabular}{l|cc}
    Datasets & with Fusion & without Fusion\\ \hline
    TW & \textbf{3.09s} \xspace [1025 rounds] & 3.55s \xspace [1489 rounds] \\
    FT & \textbf{5.64s} \xspace [5604 rounds] & 6.09s \xspace [7281 rounds]\\ 
    WB & \textbf{2.90s} \xspace [772 rounds] & 3.30s \xspace [2248 rounds]\\
    RD & \textbf{0.22s} \xspace [1069 rounds] & 0.77s \xspace [48407 rounds]\\ \hline

  \end{tabular}
  \label{table:fusion_perf_impact}
\end{table}

\begin{table}[t]
  \center \footnotesize
 \caption{Performance Impact of Eager and Lazy Bucket Updates. Lazy update for \kcore{} uses constant sum reduction optimization.}  
  \begin{tabular}{l|cc|cc}
   & \multicolumn{2}{c|}{ \kcore } & \multicolumn{2}{c}{ SSSP with \deltastepping } \\\hline
    Datasets & Eager Update & Lazy Update & Eager Update & Lazy Update\\ \hline
    LJ & 0.84  & \textbf{0.75} & \textbf{0.093} & 0.24 \\
    TW & 44.43 & \textbf{10.29} & \textbf{3.09} & 6.66 \\
    FT & 46.59 & \textbf{14.42} & \textbf{5.64} & 10.34\\
    WB & 35.58 & \textbf{12.88} & \textbf{2.90} &  7.82\\
    RD & 0.55 & \textbf{0.31} & \textbf{0.22} & 9.48\\\hline
  \end{tabular}
  \label{table:eager_lazy_perf_impact}
\end{table}

Table~\ref{table:fusion_perf_impact} shows that SSSP with bucket fusion achieves up to $3.4\times$ speedup over the version without bucket fusion on road networks, where there are a large number of rounds processing each bucket.  
Table~\ref{table:fusion_perf_impact} shows that the optimization improves running time by significantly reducing the number of rounds needed to complete the algorithm. 

Table~\ref{table:eager_lazy_perf_impact} shows the performance impact of eager versus lazy bucket updates on \kcore{} and SSSP. 
\kcore{} does a large number of redundant updates on the priority of each vertex. 
Every vertex's priority will be updated the same number of times as its out-degree. 
In this case, using the lazy bucket update approach drastically reduces the number of bucket insertions. 
Additionally, with a lazy approach, we can also buffer the priority updates and later reduce them with a histogram approach (lazy with constant sum reduction optimization). 
This histogram-based reduction avoids overhead from atomic operations. 
For SSSP there are not many redundant updates and the lazy approach introduces significant runtime overhead over the eager approach.

%% file: eval-perf-table.tex
\begin{table*}[t]\scriptsize
\centering
\tabcolsep 1.5pt
\caption{Running time (seconds) of \OG with the priority-based extension and state-of-the-art frameworks. \OG, GAPBS, Galois, and Julienne use ordered algorithms. GraphIt with no extension (unordered) and Ligra use unordered Bellman-Ford for SSSP, PPSP, wBFS, and \astar{}, and unordered \kcore{}. The fastest results are in bold.  
Graphs marked with $\dagger$ have weight distribution of $[1, \log n)$.
Road networks come with original weights. 
Other graphs have weight distribution between $[1, 1000)$.  \textbf{--} represents an algorithm not implemented in a framework and \textbf{x} represents a run that did not finish due to timeout or out-of-memory error.}
\begin{tabular}{l|ccccccc|ccccccc|ccccc}
Algorithm & \multicolumn{7}{c|}{ SSSP } & \multicolumn{7}{c|}{ PPSP } & \multicolumn{5}{c}{ wBFS } \\ \hline
Graph	&	LJ	&	OK	&	TW	&	FT	&	WB	&	GE	&	RD	&	LJ	&	OK	&	TW	&	FT	&	WB	&	GE	&	RD	&	LJ$^{\dagger}$ 	&	OK$^{\dagger}$	&	TW$^{\dagger}$	&	FT$^{\dagger}$	&	WB$^{\dagger}$	\\\hline
\OG with extension (ordered)	&	\textbf{0.093}	&	\textbf{0.106}	&	3.094	&	\textbf{5.637}	&	\textbf{2.902}	&	\textbf{0.207}	&	\textbf{0.224}	&	0.043	&	\textbf{0.061}	&	\textbf{2.597}	&	\textbf{4.063}	&	\textbf{2.473}	&	\textbf{0.049}	&	\textbf{0.045}	&	\textbf{0.072}	&	\textbf{0.104}	&	\textbf{1.822}	&	\textbf{7.563}	&	\textbf{2.129}	\\
GAPBS	&	0.1	&	0.107	&	3.547	&	6.094	&	3.304	&	0.59	&	0.765	&	\textbf{0.042}	&	0.063	&	2.707	&	4.312	&	2.628	&	0.116	&	0.112	&	0.072	&	0.107	&	1.903	&	7.879	&	2.228	\\
Galois	&	0.123	&	0.234	&	\textbf{2.93}	&	7.996	&	3.005	&	0.244	&	0.276	&	0.084	&	0.165	&	2.625	&	7.092	&	2.606	&	0.059	&	0.051	&	--	&	--	&	--	&	--	&	--	\\
Julienne	&	0.169	&	0.334	&	4.522	&	x	&	4.11	&	3.104	&	3.685	&	0.104	&	0.16	&	4.904	&	x	&	4.107	&	1.836	&	0.687	&	0.148	&	0.145	&	2.32	&	x	&	2.813	\\\hdashline
GraphIt (unordered) &0.221	&	0.479	&	6.376	&	38.458	&	8.521	&	90.524	&	122.374	&	0.221	&	0.479	&	6.376	&	38.458	&	8.521	&	90.524	&	122.374	&	0.12	&	0.198	&	2.519	&	21.77	&	3.659  \\
Ligra (unordered) &	 0.301	&	0.604	&	7.778	&	x	&	x	&	94.162	&	129.2	&	0.301	&	0.604	&	7.778	&	x	&	x	&	94.162	&	129.2	&	0.164	&	0.257	&	3.054	&	x	&	x	\\
& & & & & & & & & & & & & & & & & &  \\
Algorithm & \multicolumn{7}{c|}{ \kcore{} } & \multicolumn{7}{c|}{ Approximate Set Cover }  & \multicolumn{3}{c}{ \astar{} } & \multicolumn{2}{c}{ } \\ \hline
Graph	&	LJ	&	OK	&	TW	&	FT	&	WB	&	GE	&	RD	&	LJ	&	OK	&	TW	&	FT	&	WB	&	GE	&	RD	&	MA	&	GE	&	RD	&	\\ \hline
\OG with extension (ordered)	&	\textbf{0.745}	&	\textbf{1.634}	&	\textbf{10.294}	&	\textbf{14.423}	&	\textbf{12.876}	&	\textbf{0.173}	&	\textbf{0.305}	&	\textbf{0.494}	&	\textbf{0.564}	&	\textbf{5.299}	&	\textbf{11.499}	&	\textbf{7.57}	&	\textbf{0.545}	&	\textbf{0.859}	&	\textbf{0.010}	&	\textbf{0.060}	&	\textbf{0.075}	&	\\
GAPBS	&	--	&	--	&	--	&	--	&	--	&	--	&	--	&	--	&	--	&	--	&	--	&	--	&	--	&	--	&	0.03	&	0.142	&	0.221	&	\\
Galois	&	--	&	--	&	--	&	--	&	--	&	--	&	--	&	--	&	--	&	--	&	--	&	--	&	--	&	--	&	0.078	&	0.066	&	0.083	&	\\

Julienne	&	0.752	&	1.62	&	10.5	&	14.6	&	13.1	&	0.184	&	0.327	&	0.703	&	0.868	&	6.89	&	13.2	&	10.7	&	0.66	&	1.03	&	0.181	&	1.551	&	4.876	&	\\\hdashline
GraphIt (unordered)	&	6.131	&	8.152	&	x	&	325.286	&	x	&	0.421	&	1.757	&	--	&	--	&	--	&	--	&	--	&	--	&	--	&	0.456	&	90.524	&	122.374	&	\\
Ligra (unordered) 	&	5.99	&	8.09	&	225.102	&	324	&	x	&	0.708	&	1.76	&	--	&	--	&	--	&	--	&	--	&	--	&	--	&	0.832	&	94.162	&	129.2	&	\\\hline
\end{tabular}
\label{table:eval-perf-table}
\end{table*}

%% file: related.tex
\section{Related Work}
\label{sec:related}

\myparagraph{Parallel Graph Processing Frameworks}
There has been a significant amount of work on unordered graph processing frameworks (e.g.,~\cite{shun13ppopp-ligra, gluon2018, Zhu16gemni,Grossman2018, Yunming2017, kyrola12osdi-graphchi,Ham:2016:GHE:3195638.3195707,
    prabhakaran12atc-grace, Sakr2017,Wang:2018:LLD:3178487.3178508,
    Gonzalez2012, sundaram15vldb-graphmat, Wang:2017:GGG:3131890.3108140, GSwitch2019,
    Xu:PnP, KickStarter2017, Pai2016, graphit:2018, DBLP:journals/pvldb/SongLWGLJ18, Mukkara:18:TS,Mukkara:2019:PAS, Dhulipala:2019:LGS:3314221.3314598}, among many others).
    These frameworks do not have data structures and operators to support efficient implementations of ordered algorithms, and cannot support a wide selection of ordered graph algorithms.
A few unordered frameworks~\cite{Wang:2017:GGG:3131890.3108140,GSwitch2019, sundaram15vldb-graphmat} have the users define functions that filter out vertices to support \deltastepping{} for SSSP. This approach is not very efficient and does not generalize to other ordered algorithms.
Wonderland~\cite{Zhang:2018:WNA:3173162.3173208} uses abstraction-guided priority-based scheduling to reduce the total number of iterations for some graph algorithms.
However, it requires preprocessing and does not implement a strict ordering of
the ordered graph algorithms.
PnP~\cite{Xu:PnP} proposes direction-based optimizations for point-to-point queries, which is orthogonal to the optimizations in this paper, and can be combined together to potentially achieve even better performance.
\graphit~\cite{graphit:2018} decouples the algorithm from optimizations for unordered graph algorithms.
This paper introduces new priority-based operators to the algorithm language, proposes new optimizations for the ordered algorithms in the scheduling language, and extends the compiler to generate efficient code.


\myparagraph{Bucketing}
Bucketing is a common way to exploit parallelism and maintain ordering in ordered graph algorithms. It is expressive enough to implement many parallel ordered graph algorithms~\cite{Dhulipala:2017,DBLP:journals/corr/BeamerAP15}.
Existing frameworks support either lazy bucket update or eager bucket update approach.
GAPBS~\cite{DBLP:journals/corr/BeamerAP15} is a suite of hand-optimized C++ programs that includes SSSP using the eager bucket update approach.
Julienne~\cite{Dhulipala:2017} is a high-level programming framework that uses the lazy bucket update approach, which is efficient for applications that have a lot of redundant updates, such as $k$-Core and SetCover.
However, it is not as efficient for applications that have fewer redundant updates and less work per bucket, such as SSSP and \astar{}.
  \OG{} with the priority-based extension unifies both the eager and lazy bucket update approaches with a new programming model and compiler extensions to achieve consistent high performance.


\myparagraph{Speculative Execution}
Speculative execution can also exploit parallelism in ordered graph algorithms~\cite{Hassaan:2011:OVU:1941553.1941557,Hassaan:2015:KDG:2694344.2694363}.
This approach can potentially generate more parallelism as vertices with different priorities are executed in parallel as long as the dependencies are preserved.
This is particularly important for many discrete simulation applications that lack parallelism.
However, speculative execution in software incurs significant performance overheads as a commit queue has to be maintained, conflicts need to be detected, and values are buffered for potential rollback on conflicts.
Hardware solutions have been proposed to reduce the overheads of speculative execution~\cite{Jeffrey:Swarm,Suvinay:Fractal:2017,Jeffrey:DataCentricExectuion,Jeffrey:2018,chronos}, but it is costly to build customized hardware.
Furthermore, some ordered graph algorithms, such as approximate set cover and \kcore{}, cannot be easily expressed with speculative execution.


\myparagraph{Approximate Priority Ordering}
Some work disregard a strict priority ordering and use an approximate priority ordering~\cite{nguyen13sosp-galois, gluon2018, alistarh2015spraylist, DBLP:conf/spaa/Alistarh0KLN18}.
This approach uses several ``relaxed" priority queues in parallel to maintain local priority ordering.
However, it does not synchronize globally among the different priority queues.
To the best of our knowledge and from communications with the developers, Galois~\cite{nguyen13sosp-galois, gluon2018} does not currently support strict priority ordering and only supports an approximate ordering.
Galois~\cite{nguyen13sosp-galois} provides an ordered list abstraction, which does not explicitly synchronize after each priority. 
As a result, it is hard to implement algorithms that require explicit synchronization, such as \kcore{}.
Galois also require users to handle atomic synchronizations for correctness.
This approach cannot implement certain ordered algorithms that require strict priority ordering, such as work-efficient \kcore{} decomposition and SetCover. 
D-galois~\cite{Dathathri:2019:PSR:3297858.3304056} implements \kcore{} for only a specific $k$, whereas \OG{}'s \kcore{} finds \emph{all} cores. 


\myparagraph{Synchronization Relaxation}
There has been a number of frameworks that relax synchronizations in graph algorithms for better performance while preserving correctness~\cite{Harshvardhan:2014:KNA, Vora:2014:AEA, Ben-Nun:2017:GAM}. 
Compared to existing synchronization relaxation work, bucket fusion in our \extension is more restricted on synchronization relaxation. 
The synchronization between rounds can be removed only when the vertices processed in the next round has the same priority as vertices processed in the current round. 
This way, we ensure no priority inversion happens.

%% file: artifact.tex
\appendix 
\section{Artifact Evaluation Information}

\begin{itemize}
  \item {\bf Algorithms:} SSSP with \deltastepping{}, PPSP, wBFS, \astar{}, \kcore, and Approximate Set Cover 
  \item {\bf Compilation:} C++ compiler with C++14 support, Cilk Plus and OpenMP
  \item {\bf Binary:} Compiled C++ code
  \item {\bf Data set:} Social, Web, and Road graphs
  \item {\bf Run-time environment:} Ubuntu 11.04
  \item {\bf Hardware:} 2-socket Intel Xeon E5-2695 v3 CPUs with Transparent Huge Pages enabled
  \item {\bf Publicly available?} Yes
  \item {\bf Code licenses (if publicly available)?} MIT License
\end{itemize}

The detailed instructions to evaluate the artifact are available at \url{https://github.com/GraphIt-DSL/graphit/blob/master/graphit_eval/priority_graph_cgo2020_eval/readme.md}.

The evaluation in the link first demonstrates how SSSP with $\Delta$-stepping with different schedules are compiled to C++ programs (Figure~\ref{fig:sssp_different_schedules}). Then we provide instructions on how to run different algorithms on small graphs serially. Finally, there is an optional part that replicates the parallel performance on a more powerful 2-socket machines for LiveJournal, Twitter, and RoadUSA graphs (Table~\ref{table:eval-perf-table}). 

%% file: ack.tex
\section*{Acknowledgments}
We thank Maleen Abeydeera for help with \astar{} 
and Mark Jeffrey for helpful comments. 
This research was supported by DOE Early Career Award
\#DE-SC0018947, NSF CAREER Award \#CCF-1845763, MIT Research Support Committee Award, DARPA SDH Award
\#HR0011-18-3-0007, Applications Driving Architectures (ADA)
Research Center, a JUMP Center co-sponsored by SRC and DARPA, 
Toyota Research Institute, DoE Exascale award \#DE-SC0008923,
DARPA D3M Award \#FA8750-17-2-0126.